\documentclass{article}

\usepackage[english]{babel}
\usepackage{csquotes}

\usepackage[letterpaper,top=2cm,bottom=2cm,left=3cm,right=3cm,marginparwidth=1.75cm]{geometry}

\usepackage{amsmath}
\usepackage{graphicx}
\usepackage[colorlinks=true, allcolors=blue]{hyperref}
\usepackage{float}
\usepackage[
    backend=biber,
    style=authoryear,
    citestyle=authoryear,
    sorting=nyt,
    giveninits=true,
    uniquename=init,
    maxcitenames=2,
    mincitenames=1
]{biblatex}

\usepackage{multirow}
\usepackage{authblk}

\title{A Conservative Darcy–Brinkman–Stokes Formulation for Digital Rock Flow Simulation with Embedded Second-Brinkman Effects}

\author{\small
Gospel E. Stewart\thanks{Corresponding author:
\href{mailto:ges2000@hw.ac.uk}{ges2000@hw.ac.uk}},
Hannah P. Menke, David Egya, Jacqueline Mifsud,
and Julien Maes}

\affil{Institute for GeoEnergy Engineering, Heriot-Watt University, Edinburgh, United Kingdom}

\begin{document}
\maketitle

\begin{abstract} 
 Accurate pore-scale direct numerical simulation (DNS) typically requires boundary-conforming meshes on fully resolved geometries, which is memory-intensive and often becomes impractical when modelling complex rock geometries. In contrast, the Darcy–Brinkman–Stokes (DBS) framework enables simulations to be performed directly on voxelized micro-CT images. However, the commonly used simplified DBS formulation (DBS1) neglects the second Brinkman term (SBT) and the porosity-gradient viscous-diffusion term, leading to errors in flow-field characterisation and permeability prediction. Although it is well established that the SBT is essential for capturing viscous boundary effects, it is often assumed that its inclusion requires separate discretisation. In this work, we show that no separate discretisation of the SBT is necessary: writing the DBS equation in its conservative form and applying a standard central differencing scheme in a finite-volume discretisation naturally retains the full DBS formulation. We refer to this formulation as DBS2, and validate its implementation in the GeoChemFoam framework (our open-source computational fluid dynamics simulator based on OpenFOAM) using three benchmark cases: (i)  microchannel flow, (ii) a 2D heterogeneous micromodel, and (iii) a 3D micro-CT image of Bentheimer sandstone. Simulation results show that the DBS2 formulation reproduces the DNS velocity field exactly in the microchannel and matches DNS permeability to within 1.5–3.8\% in the micromodel and Bentheimer cases at matched spatial resolution. In contrast, DBS1 systematically overpredicts permeability (14.1 to 37.9\% in our cases) under the same discretization. Notably, the conservative treatment of the viscous stress preserves the computational efficiency of DBS1 while achieving DNS-level accuracy.
\end{abstract}

\noindent\textbf{Keywords:}
Darcy--Brinkman--Stokes; second Brinkman term; pore-scale modelling; digital rock physics; finite-volume method; permeability.
\par\medskip

\subsection*{Article Highlights}
\begin{itemize}
    \item Viscous effects near solid boundaries are captured
    without a separate correction term.

    \item The revised model reproduces reference flow
    predictions in a straight microchannel.

    \item Permeability errors are below 4\% in a heterogeneous
    micromodel and a sandstone image.
\end{itemize}

\section{Introduction}

Fluid flow in porous media governs a wide variety of natural and engineered processes, from the migration of hydrocarbons in subsurface reservoirs \parencite{Bear1972} to the transport of reactants in batteries and fuel cells \parencite{Larminie2003}. It also plays a central role in energy transition technologies, such as carbon capture and storage \parencite{Szulczewski2012}, hydrogen storage \parencite{Heinemann2021}, and geothermal energy \parencite{Pruess_1988_modelingOfGeothermalReservoirs, OSullivan_Pruess_Lippmann_2000_stateOfPractice}. Porous media exhibit complex physics across multiple length scales, from nanometer-scale pores to kilometer-scale reservoirs, necessitating different modelling approaches to capture the dominant mechanisms at each scale. At the macroscopic scale, models integrate geometric, material, and thermophysical data to simulate long-term fluid flow and transport in porous systems. The continuum scale provides an intermediate framework, treating porous media as homogenised continua via averaged laws such as Darcy’s law, but is often constrained by the representativeness of effective properties. 

At the pore-scale, direct geometric and physical resolution of the pore structure provides fundamental insights into flow and transport phenomena within the porous media. Recent advances in high-resolution imaging techniques, particularly X-ray microtomography (micro-CT), coupled with ongoing developments in Digital Rock Physics (DRP) and enhanced computational capabilities, have transformed pore-scale modelling from a conceptual framework into a predictive tool capable of resolving complex flow and transport processes. Three-dimensional (3D) images of rock and other microstructures can now be routinely acquired at micrometer-scale and higher resolution, enabling the direct simulation of flow within real pore spaces \parencite{Berg2017}. These developments have established pore-scale modelling as a powerful complement to laboratory experiments, enabling accurate prediction of critical transport properties, including absolute and relative permeability \parencite{khan2012simulation, andra2013digital}, longitudinal and transverse dispersion coefficients \parencite{bijeljic2004pore, maes2024dispersivity, ortega-ramirez2021effect}, and fundamental constitutive relationships for continuum-scale modelling, such as porosity-permeability correlations \parencite{schepp2020digital, osorno2015finite}.

Traditional approaches to pore-scale modelling include Pore Network Modelling (PNM) and Direct Numerical Simulation.  PNM simplifies the porous medium into a network of interconnected pores and throats, allowing efficient simulation of flow and transport processes \parencite{blunt1992}. While PNM is computationally less demanding than DNS and can handle large-scale systems, it lacks the detailed resolution of the actual pore structure, which can lead to potential inaccuracies in capturing fine-scale flow dynamics and transport phenomena. An alternative strategy is to use DNS, which solves the governing conservation equations for detailed pore geometries reconstructed from imaging data. This approach encompasses methods such as the Lattice Boltzmann Method (LBM) \parencite{Ramstad2019} and finite-volume- and finite-element-based methods \parencite{Zaretskiy2010}.

DNS, based on the Navier-Stokes equations, is widely regarded as the benchmark method in pore-scale modelling due to its ability to deliver highly detailed and accurate simulations of fluid flow and transport. DNS directly resolves the fluid dynamics at the pore scale, capturing intricate flow patterns and interactions within the pore space. This method is particularly well-suited for investigating complex fluid behaviour and reactive flows, making it a preferred choice when accuracy is required \parencite{golparvar2018}. The standard DNS approach relies on an explicit void–solid representation of the pore space, requiring accurate binary segmentation of imaging data \parencite{kang2019}. It also typically requires boundary-fitted meshes to resolve the solid–liquid interface, which can be prohibitively memory-intensive for heterogeneous natural porous media. 

The DBS framework provides a practical alternative by allowing simulations directly on voxel-based micro-CT images \parencite{soulaine2016micro}. In this one-domain approach, porosity and permeability are modelled as continuous fields, enabling a single set of partial differential equations to describe transport across solid and free-fluid regions. In the free-fluid region, the equations reduce to Navier–Stokes flow, while solid voxels are represented as regions of very low porosity and permeability, allowing the DBS model to act as an approximate immersed boundary method (IBM), where the solid-fluid interface is handled implicitly through penalisation without explicit boundary treatment \parencite{Liu2007BrinkmanPenalization}. This formulation has been widely applied to reactive flow  \parencite{Yang2021, soulaine2017, maes2022}, deformable porous media \parencite{carrillo2019}, heat transfer \parencite{maes2022geom} and multiphase systems \parencite{HORGUE2014}. The DBS framework can be extended to multiscale systems \parencite{soulaine2016micro, Soulaine2024, Carrillo2020}, enabling simulations across heterogeneous domains with minimal additional computational cost \parencite{yang2021DBSproblems}. 

The DBS equations are commonly derived by applying volume-averaging techniques to the local conservation laws \parencite{Whitaker1999}. This procedure yields an averaged momentum equation containing three contributions: a Darcy drag term representing momentum loss due to the solid matrix, a Brinkman correction accounting for viscous diffusion within the pore space, and a higher-order contribution known as the second Brinkman term (SBT), which arises from pore-scale velocity fluctuations. In many studies, the SBT is neglected under strict scale separation assumptions, an approximation supported by early theoretical analyses \parencite{Quintard1994}. However, subsequent work has shown that this simplification may be limiting, particularly at fluid–porous interface regions where scale separation breaks down. The foundational studies of \textcite{OchoaTapia1995a, OchoaTapia1995b} demonstrated that omitting the SBT leads to systematic errors in interfacial velocity predictions. Similar conclusions have since been reported in multicomponent solidification \parencite{BousquetMelou2002} and more recent investigations of porous media flow by \textcite{HernandezRodriguez2022, Angot2021}.

The explicit implementation of the SBT, as demonstrated in a recent study by \textcite{Machado2025Correcting}, discretised the SBT gradients using a Gaussian scheme with harmonic interpolation. This ensures that gradient terms are handled more accurately, especially when the mesh is coarse. These findings highlight two unresolved issues in DBS modelling: the appropriate definition of the effective viscosity appearing in the Brinkman terms, and whether inclusion of the SBT is necessary to accurately capture interfacial and multiscale behaviour.

In this study, we revisit the DBS momentum formulation from the perspectives of consistency and conservation. We adopt and validate a conservative Brinkman formulation that yields a physically consistent definition of the effective viscosity while naturally retaining higher-order contributions arising from volume averaging. The resulting model preserves the structural simplicity of the classical DBS equation, while addressing known deficiencies in simplified formulations. By comparing this approach against both simplified DBS models and DNS benchmarks, we quantify its impact on predicted flow velocities and macroscopic properties such as permeability. The following section introduces the governing equations and their numerical implementation, and then presents results for a hierarchy of test cases ranging from a simple microchannel with a solid boundary to a 3D image-based porous medium.

\subsection{Governing Equations}
\subsubsection{Navier Stokes Equations}
For an incompressible, Newtonian fluid flow, the Navier-Stokes mass conservation and momentum equations for a system can be written as
\begin{equation}
\nabla \cdot \mathbf{u} = 0,
\label{eq.continuity}
\end{equation}
\begin{equation}
\frac{\partial \mathbf{u}}{\partial t} + \nabla \cdot (\mathbf{u}\mathbf{u}) = -\nabla p + \nabla \cdot (\nu \nabla \mathbf{u}) + \mathbf{f}_b,
\label{eq.momentum}
\end{equation}
where \(\mathbf{u}\) (m/s) is the velocity field, \(p\) (m$^2$/s$^2$) is the kinematic pressure, \(\mathbf{f}_b\) (m/s$^2$) represents other body forces (e.g., gravity), \(\nu \) (m$^2$/s) is the kinematic viscosity of the fluid, and \(t\) (s) is time. For a steady-state flow at the pore scale with low Reynolds number ($\text{Re} \ll 1$) and no additional body forces, the momentum equation above reduces to the Stokes equation of the form
\begin{equation}
\nu\nabla^2 {\mathbf{u}} = \nabla p.
\label{eq:stokes}
\end{equation}
The interfacial boundary conditions for the system are:
\begin{align*}
\mathbf{u} &= 0 \quad \text{(no-slip at the boundaries)}, \\
\nabla p \cdot \mathbf{n} &= 0 \quad \text{(where $\mathbf{n}$ is the normal to the interface).}
\end{align*}
Note that the pressure condition is not an independent physical boundary condition of the incompressible Navier–Stokes equations; rather, it is commonly imposed as a numerical closure for the pressure Poisson equation arising in pressure–velocity coupling schemes \parencite{2006-Sani,PATANKAR19721787}.

\subsubsection{Darcy-Brinkmann-Stokes Equations}

The one-domain Darcy--Brinkman--Stokes (DBS) momentum equation governing 
steady, incompressible, non-inertial flow in a heterogeneous porous medium 
may be written in expanded form as~\parencite{2017-Angot}
\begin{align}
  0
  = -\nabla\bar{p}
  + \frac{\nu}{\varepsilon}\nabla^{2}\bar{\mathbf{u}}
  - \frac{\nu}{\varepsilon}\,\nabla\varepsilon
      \cdot\nabla\!\left(\frac{\bar{\mathbf{u}}}{\varepsilon}\right)
  - \nu\mathbf{K}_{d}^{-1}\bar{\mathbf{u}},
  \label{eq:generalisedDBS}
\end{align}
where $\bar{p}$ is the volume-averaged pressure, $\bar{\mathbf{u}}$ is the 
superficial (Darcy) velocity, $\nu$ is the kinematic viscosity, $\varepsilon$ 
is the local porosity, and $\mathbf{K}_{d}$ is the permeability tensor 
associated with the Darcy resistance.
The four terms on the right-hand side of Eq.~\eqref{eq:generalisedDBS} 
represent, respectively, the pressure gradient driving the flow, viscous 
diffusion of momentum scaled by the inverse porosity, an additional viscous 
contribution that arises from spatial gradients of porosity, and the Darcy 
drag exerted by the solid region.
The third term — hereafter called the second Brinkman term (SBT) — is 
active only where $\nabla\varepsilon \neq 0$, that is, within the 
interfacial transition region between the free-fluid and porous domains; 
and it vanishes identically in any homogeneous region.

Following the conservative reformulation of \textcite{2017-Angot}, Eq.~\eqref{eq:generalisedDBS} can be rewritten as
\begin{align}
  0
  = -\nabla\bar{p}
  + \nabla\cdot\!\left(\frac{\nu}{\varepsilon}\nabla\bar{\mathbf{u}}\right)
  - \nu\mathbf{K}^{-1}\bar{\mathbf{u}},
  \label{eq:compactDBS}
\end{align}
where the permeability tensor appearing in the drag term is modified 
relative to $\mathbf{K}_{d}$ according to
\begin{align}
  \mathbf{K}^{-1}
  =
  \mathbf{K}_{d}^{-1}
  -\frac{|\nabla\varepsilon|^{2}}{\varepsilon^{3}}\,\mathbf{I}.
  \label{eq:modifiedPermeability}
\end{align}
The isotropic correction $|\nabla\varepsilon|^{2}/\varepsilon^{3}$ 
in Eq.~\eqref{eq:modifiedPermeability} is positive definite and as shown in Appendix~A of \textcite{2017-Angot}, $\mathbf{K}$ remains symmetric and positive definite, provided that 
$\varepsilon^{-3}|\nabla\varepsilon|^{2}\mathbf{K}_{d} \ll \mathbf{I}$, 
a condition that is satisfied whenever porosity gradients are small relative to the characteristic length scale of porosity variation. The algebraic equivalence of Eqs.~\eqref{eq:generalisedDBS} and~\eqref{eq:compactDBS} is established by expanding the divergence-form 
viscous operator.

Consequently, within the conservative formulation:
(i)~the SBT is already encoded inside the divergence-form viscous operator 
and requires no separate discretisation;
(ii)~the residual $|\nabla\varepsilon|^{2}/\varepsilon^{3}$ contribution 
is absorbed into the modified permeability $\mathbf{K}^{-1}$ rather than 
appearing as an independent term.

The relevance of this result to the present voxel-based penalisation 
framework is as follows. In GeoChemFoam, the porosity field takes the discrete values 
$\varepsilon = 1$ in pore voxels and $\varepsilon = \varepsilon_{\min}=1 \times 10^{-4}$ in solid voxels; non-zero porosity gradients are therefore confined to the interfacial layer between the fluid and solid regions. In this setting, the isotropic correction $|\nabla\varepsilon|^{2}/\varepsilon^{3}$ 
acts exclusively as a local additional resistance concentrated in 
interfacial and solid cells. It can therefore be incorporated into the penalisation of the permeability, which is already prescribed in those cells to enforce the no-slip condition on solid boundaries; and therefore, no separate explicit treatment of this term is needed.

From a numerical standpoint, the conservative form is also practically advantageous because it preserves the standard diffusion--drag operator structure of the DBS equation, allowing the viscous contribution to be handled with the finite-volume discretisation schemes that are native to 
OpenFOAM.
In the present work, this is implemented by interpolating the effective diffusivity $\nu/\varepsilon$ to cell faces, yielding the face-centred coefficient $\nu_{f}/\varepsilon_{f}$ that is used to compute the 
viscous flux across each pore--solid transition. Since $(\varepsilon_f \neq 0)$ at pore–solid faces, the division-by-zero issue is confined to cells located entirely within the solid phase, where $(\varepsilon \approx 0)$.

\subsubsection{Simplified DBS Equation}
The classical DBS equation, as adopted in many studies relating to this subject, is a subset of the full DBS where the SBT is neglected and can be written as: 

\begin{align}
0
  &= -\,\nabla \bar{p}
     + \frac{\nu}{\varepsilon} \nabla^{2}\mathbf{\bar{u}}
     - \nu \mathbf{K}^{-1}\mathbf{\bar{u}}
     %+ \mathbf{F},
  \label{eq:dbs_momentum}
  \\[6pt]
  \nabla\cdot\mathbf{\bar{u}} &= 0.
  \label{eq:dbs_continuity}
\end{align}

\noindent
This equation incorporates several physical phenomena through its constituent terms \parencite{Auriault2009}:

\begin{itemize}
    \item \textbf{Pressure gradient:} 
    The term 
    \( -\nabla \bar{p} \)
    drives the flow due to pressure differences across the domain.

    \item \textbf{Brinkman viscous term:} 
    The term 
    \( \frac{\nu}{\varepsilon} \nabla^2 \mathbf{\bar{u}} \)
    represents viscous diffusion, analogous to the Stokes equation in the presence of a local porosity $\varepsilon$. 

    \item \textbf{Darcy drag term:} 
    The term 
    \(- \nu \mathbf{K}^{-1}\mathbf{\bar{u}}\)
    represents the linear resistance to flow imposed by the solid boundaries, consistent with Darcy’s law.

\end{itemize}

\subsection{Permeability in the DBS formulation}

In the Darcy--Brinkman--Stokes (DBS) formulation, the drag term is
controlled by the permeability tensor, $\mathbf{K}$, which represents the
resistance of each computational cell to fluid motion. In this work, we assume
an isotropic permeability, such that $\mathbf{K}=K\mathbf{I}$, and prescribe the
scalar permeability $K$ using a Kozeny--Carman-type relation \parencite{Kozeny1927, Carman1937}:
\begin{equation}
K = K_f \frac{\varepsilon^3}{(1-\varepsilon)^2},
\label{eq:KC_modified}
\end{equation}
where $K_f$ is a fitted permeability scale. The parameter $K_f$ is chosen numerically so that the permeability
assigned to the solid region is sufficiently small, typically of the order of $K_f \sim 10^{-21}\,\mathrm{m^2}$. This ensures strong momentum penalisation in solid cells. The corresponding inverse permeability $K^{-1}$ in the pore space where $\phi=1$, vanishes, i.e.
\begin{equation}
K^{-1}=0,
\end{equation}
and hence no artificial drag force is applied in the fluid region. Conversely, in the solid region, where $\phi \ll 1$, $K^{-1}$ becomes large, leading to a strong damping of the velocity and approximating the no-slip condition at the fluid-solid interface.

\section{Numerical Implementation in GeoChemFoam}

GeoChemFoam is our open-source pore-scale numerical simulator based on OpenFOAM \parencite{GeoChemFoam2024} and employs a cell-centred finite-volume method (FVM), where primary unknowns are stored at cell centres, and fluxes are evaluated at control-volume faces. In the implementation of the DBS method, the governing momentum equation is discretised in its conservative form, with contributions 
from transient, convective, diffusive, and resistive terms. 
In this framework, convective fluxes are computed via face interpolation of cell-centred velocities, while diffusive fluxes are represented through face-based gradients. 
Darcy drag terms are incorporated as source contributions proportional to the local permeability tensor. 

\subsection{Implicit Capture of the Second-Order Brinkman Term}
A central question addressed in this work is whether explicit discretisation of the second Brinkman term is necessary for accurate pore-scale simulations.
An essential feature of the present implementation of the DBS equation in GeoChemFoam is the implicit treatment of the viscous term through an effective diffusion coefficient defined at cell faces,
\begin{equation}
\nu_{\mathrm{eff},f} = \frac{\nu_f}{\varepsilon_f},
\end{equation}
where $\nu_f$ and $\varepsilon_f$ denote face-centred viscosity and porosity, respectively. Face values are obtained by linear interpolation of neighbouring cell-centred quantities, with non-orthogonal corrections applied to ensure accuracy on general unstructured meshes. In GeoChemFoam, this viscous contribution is implemented using a variable-coefficient Laplacian operator,
\begin{equation}
\texttt{fvm::laplacian(fvc::interpolate(nu)/fvc::interpolate(eps), U)}.
\end{equation}

Although written in conservative form, this discretisation naturally captures both the first and second Brinkman viscous terms through the divergence of the effective diffusion coefficient, without requiring explicit storage of porosity gradients or specialised discretisation schemes. The implicit treatment is mathematically equivalent to the corresponding explicit formulation while maintaining improved numerical robustness and computational efficiency. In classical DBS formulations, expansion of the viscous term leads to factors of $1/\varepsilon$ multiplied by the full divergence operator, which can result in numerical stiffness or breakdown as $\varepsilon \rightarrow 0$ in solid regions. In contrast, the conservative formulation places the porosity dependence inside the divergence operator, such that velocity gradients naturally vanish in the solid domain where $\bar{\mathbf{u}} \rightarrow 0$. As a result, the formulation remains well-defined even for $\varepsilon \approx 0$, avoiding singular behaviour while consistently enforcing impermeability of the solid phase. These properties make the conservative form particularly attractive for pore-scale simulations involving sharp fluid--solid interfaces and highly heterogeneous porous structures. In the following sections, we present validation and application test cases demonstrating the performance of the DBS implementation in GeoChemFoam.

\section{Model Validation}
To validate and establish the applicability of the implementation presented in the previous section, we select specific cases of different sizes and heterogeneity. These cases are designed to assess the method’s accuracy, robustness, and computational behaviour as geometric and structural complexity increases. The results are analysed to compare the first- and second-order DBS with those from the conventional DNS (Stokes-based) model. For simplicity, in the remainder of this analysis, we denote DNS as cases with no solid in the numerical domain and all solutions resolved to Stokes flow; DBS1 as the classical implementation; and DBS2 as our new implementation of the conservative second-order DBS equation in GeoChemFoam. 

A typical setup for most test cases in this study involves obtaining a digital model of the domain—a geometric image of the model—followed by meshing the domain and conducting numerical flow simulations on the resulting mesh. Some images used in this analysis are generated using image-processing software (e.g., Fiji) or Python scripts, or acquired via X-ray tomography.  
Mesh construction is a crucial aspect of this analysis. Research indicates that different pore-space mesh construction methods can influence parameter estimations \parencite{Guibert2015Permeability}. The two main meshing techniques commonly used in pore-scale modelling are voxelized meshes and boundary-fitted ones. Past studies have shown that both approaches can produce similar permeability results when the image’s grid resolution is sufficiently high (usually in fully resolved images) \parencite{Blunt2013_PermeabilityResolution, Song2019_ComparativePermeability}. This study uses voxel-based pore-space meshes, where each square or cubic voxel directly represents a phase or region within the modelling framework. In DBS cases (DBS1 and DBS2), this involves meshing the entire model, including the solid, with Cartesian grids. All prescribed pressure values are kinematic pressures, $p=P/\rho$, and are reported in $\mathrm{m^2/s^2}$, where $P$ is the physical pressure and $\rho$ is the fluid density. 

\subsection {Horizontal Fracture \label{subsec:2D_Fracture}}
\begin{figure}[H]
  \centering
  \includegraphics[width=0.9\textwidth]{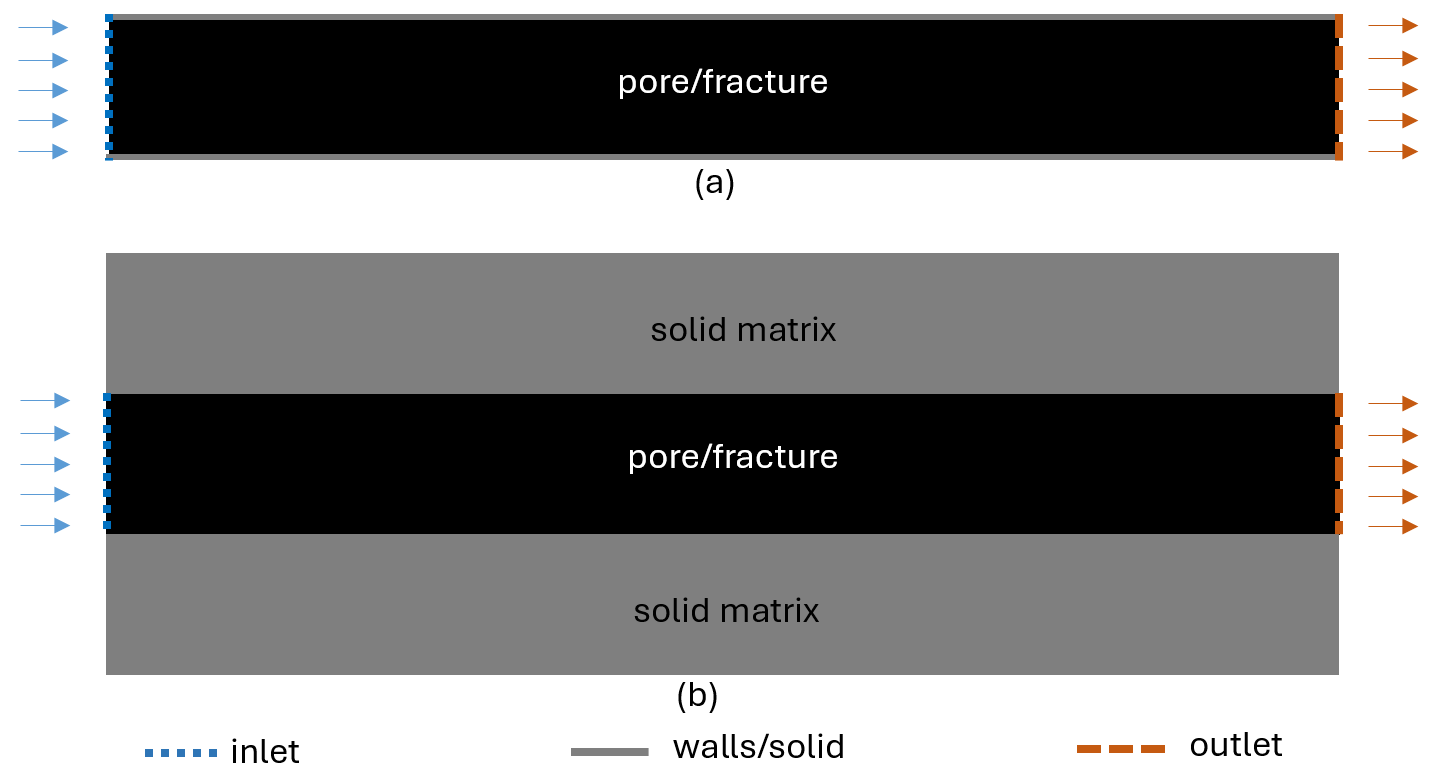}
  \caption{Model setup for the microchannel simulation (a) DNS (b) micro-continuum (DBS1 and DBS2).}
  \label{fig:micromodel_schematic}
\end{figure}

The first validation case examined involves flow through a horizontal fracture with solid walls. This setup was chosen because it is mathematically equivalent to pressure-driven flow between two parallel plates (microchannel), for which an analytical solution exists. Consequently, it provides a straightforward benchmark for evaluating the accuracy of the numerical implementation, enabling a direct comparison between simulated and theoretical velocity and permeability values. Figure~\ref{fig:micromodel_schematic} illustrates the horizontal microchannel, measuring 500 micrometres in length and 50 micrometres in width. This is the configuration used for the DNS simulation, where there is no solid in the numerical domain (Figure~\ref{fig:micromodel_schematic}a). For the DBS methods, the model includes solid boundaries at the top and bottom of the channel, each 50 micrometres thick (Figure~\ref{fig:micromodel_schematic}b).

This model configuration is specified in the OpenFOAM \textit{blockMeshDict} file, which serves as the model creation dictionary. The resulting 2D model is meshed using the blockMesh utility in OpenFOAM to generate a uniform Cartesian grid. In the traditional DNS case, this involves meshing only the channel pore space, whereas in the DBS cases, it means meshing the entire domain, including the solid region. In the DBS formulation, the solid phase is defined as a matrix with very low porosity and permeability. The porosity of the fluid cells is set to 1 in all cases, whereas the solid cells are set to 1e-4 in the DBS simulations, the minimum allowed value that guarantees convergence, particularly with the classical DBS implementation. To assign material properties, such as porosity, to the model, we use the setFields utility in OpenFOAM to assign values to the fields in the numerical domain.
The flow simulation for this model was performed using two distinct formulations of the DBS method, DBS1 and DBS2 introduced earlier. The results are compared with those from the conventional single-phase solver in OpenFOAM (simpleFoam) and with the analytical solution for a fully developed velocity profile. 

\begin{figure}[H]
  \centering
  \includegraphics[width=0.9\textwidth]{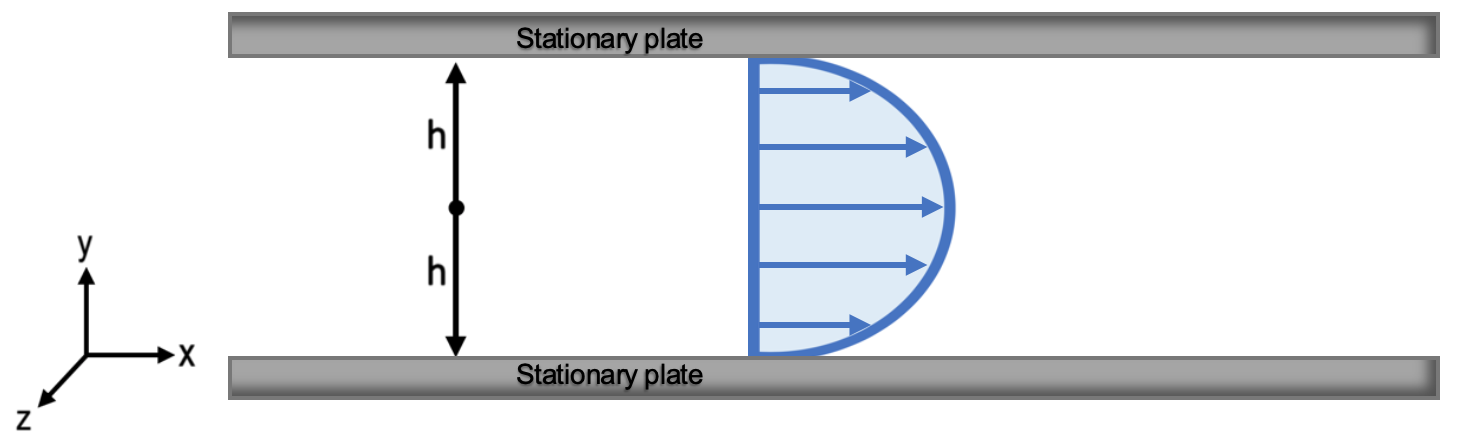}
  \caption{Schematic illustration of fully developed laminar flow between two parallel plates.}
  \label{fig:flowinparallelplates_schematic}
\end{figure}

For the laminar flow between two parallel plates shown in Figure \ref{fig:flowinparallelplates_schematic}, the velocity distribution can be expressed as
\begin{equation}
u_{x}(y) \;=\; u_{\text{max}} \left( 1 - \left( \frac{y}{h} \right)^{2} \right),
\end{equation}
where:
\begin{itemize}
\item $u_{x}(y)$ is the velocity at a distance $y$ from the centerline, 
$u_{\text{max}}$ is the maximum velocity at the centerline ($y=0$)
\item $h$ is half the distance between the plates. The velocity is zero at the plates, 
($y = \pm h$), due to the no-slip boundary condition. 
\end{itemize}

\subsubsection{Permeability Estimation Model}

In the validation steps of this study, we also compare the predicted absolute 
permeabilities obtained from different methods. For fully developed, steady flow between two parallel solid plates separated by a distance $2h$, the analytical
permeability is
\begin{equation}
K_x = \frac{h^2}{3}.
\end{equation}

\subsubsection{Mesh Resolution}
Mesh convergence testing provides confidence that the results accurately reflect the underlying physics of the flow rather than numerical discretisation artefacts. In this analysis, we examine different mesh resolutions and compare the convergence of the various methods.  Table~\ref{tab:mesh_refinement} shows the multiple levels of refinement and the corresponding total mesh cells for the DNS and DBS methods.
\begin{table}[H]
\centering
\caption{Mesh refinement parameters and corresponding cell counts.}
\label{tab:mesh_refinement}
\begin{tabular}{cccc}
\hline
\multirow{2}{*}{Refinement Factor} & \multirow{2}{*}{Grid Size (\(\mu\)m)} & \multicolumn{2}{c}{Number of Cells} \\
\cline{3-4}
& & DNS (500\(\mu\)m ×50 \(\mu\)m) & DBS1 and DBS2 (500\(\mu\)m ×150\(\mu\)m)\\
\hline
1  & 10.0  & 250    & 750     \\
2  & 5.0   & 1,000  & 3,000   \\
4  & 2.5   & 4,000  & 12,000  \\
8  & 1.25  & 16,000 & 48,000  \\
16 & 0.625 & 64,000 & 192,000 \\
\hline
\end{tabular}
\end{table}

\subsubsection{Numerical Simulation Setup}

\subsubsection*{Simulation Properties}
The test case is formulated as a single-phase flow through a horizontal microchannel with a kinematic pressure differential imposed across the model. Table~\ref{tab:openfoam_parameters} shows a summary of the properties of the solid and fluid used for this simulation. 

\begin{table}[H]
\centering
\caption{Simulation parameters for microchannel case.}
\begin{tabular}{l c}
\hline
\textbf{Property} & \textbf{Value} \\
\hline
Inlet Pressure & $1\times10^{-4}~\mathrm{m^2/s^2}$ \\
Outlet Pressure & $0~\mathrm{m^2/s^2}$ \\
Fluid Viscosity & $1 \times 10^{-6}~\mathrm{m^2/s}$ \\
\hline
\end{tabular}
\label{tab:openfoam_parameters}
\end{table}

\subsubsection*{Boundary Conditions}

The implemented boundary conditions are summarised in Table~\ref{tab:bc_microchannel}. This setup ensures proper representation of laminar flow in a 2D microchannel while maintaining consistent boundary treatments for both fluid and solid regions.

\begin{table}[H]
\centering
\caption{Boundary conditions for 2D Microchannel Simulation in OpenFOAM\textsuperscript{\textregistered}}
\begin{tabular}{l l l}
\hline
\textbf{Boundary} & \textbf{Field} & \textbf{Boundary Condition} \\
\hline
Inlet & Pressure & fixedValue  \\
      & Velocity & zeroGradient \\
Outlet & Pressure & fixedValue  \\
       & Velocity & zeroGradient \\
Solid walls & Velocity & noSlip \\
Front and Back & All fields & empty \\
\hline
\end{tabular}
\label{tab:bc_microchannel}
\end{table}

\subsubsection{Results and Discussions}
\subsubsection*{Velocity Distribution and Profile}
\begin{figure}[H]
  \centering
  \includegraphics[width=1.1\textwidth]{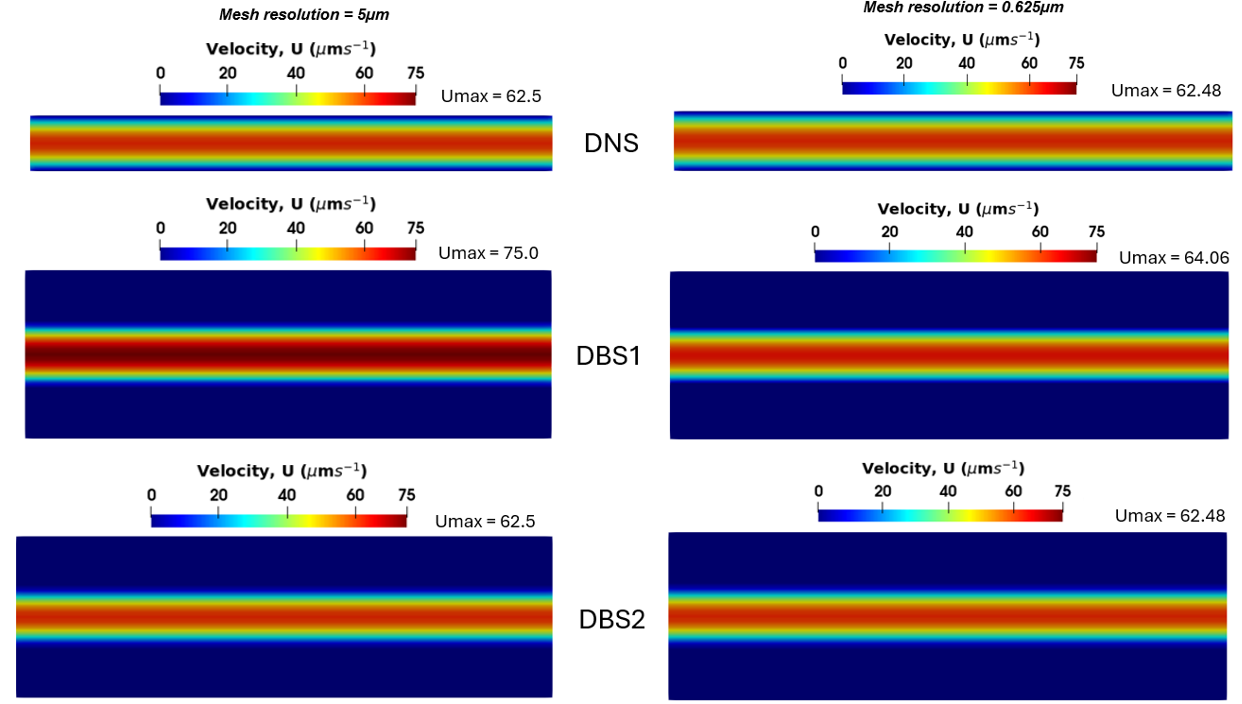}
  \caption{Simulation results showing the velocity distribution in the microchannel for two different mesh resolutions (coarse and fine).}
  \label{fig:velocityField}
\end{figure}

Figure \ref{fig:velocityField} presents a comparison of velocity fields obtained using the different methods. The simulations show velocity distributions in the microchannel, resolved with a coarse (5$\mu$m) and a fine (0.625$\mu$m) mesh. 

These results demonstrate that the second-order DBS implementation (DBS2) agrees well with the DNS method, even at coarse-mesh resolutions, compared with the first-order implementation (DBS1). The accuracy of the DBS1 improves with finer mesh resolution, approaching the DNS results.

\begin{figure}[H]
  \centering
  \includegraphics[width=0.8\textwidth]{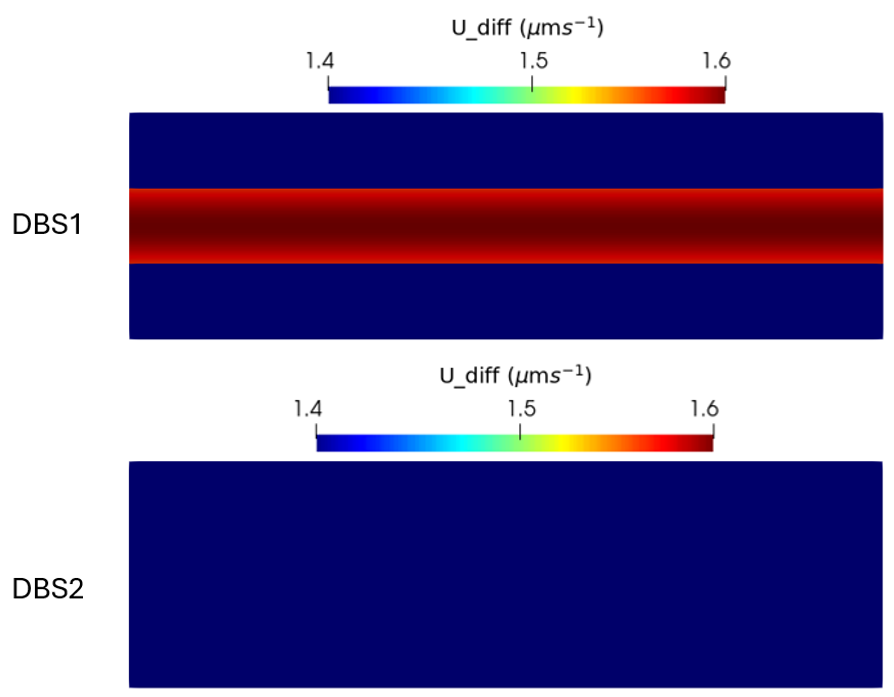}
  \caption{Velocity difference map in the pore-space for DBS1 and DBS2 in comparison to the DNS method ($0.625\mu m$).}
  \label{fig:channel_diff_map}
\end{figure}

Figure~\ref{fig:channel_diff_map} illustrates the spatial distribution of the velocity differences obtained using the DBS1 and DBS2 methods relative to the reference DNS solution. The colour scale represents the magnitude of the velocity difference, $U_{diff}$ (m/s). The DBS2 result exhibits a uniform field with negligible deviation from the DNS velocities. In contrast, the DBS1 field shows noticeable variations, reflecting localised discrepancies in the predicted flow velocities.

\begin{figure}[H]
  \centering
  \includegraphics[width=0.7\textwidth]{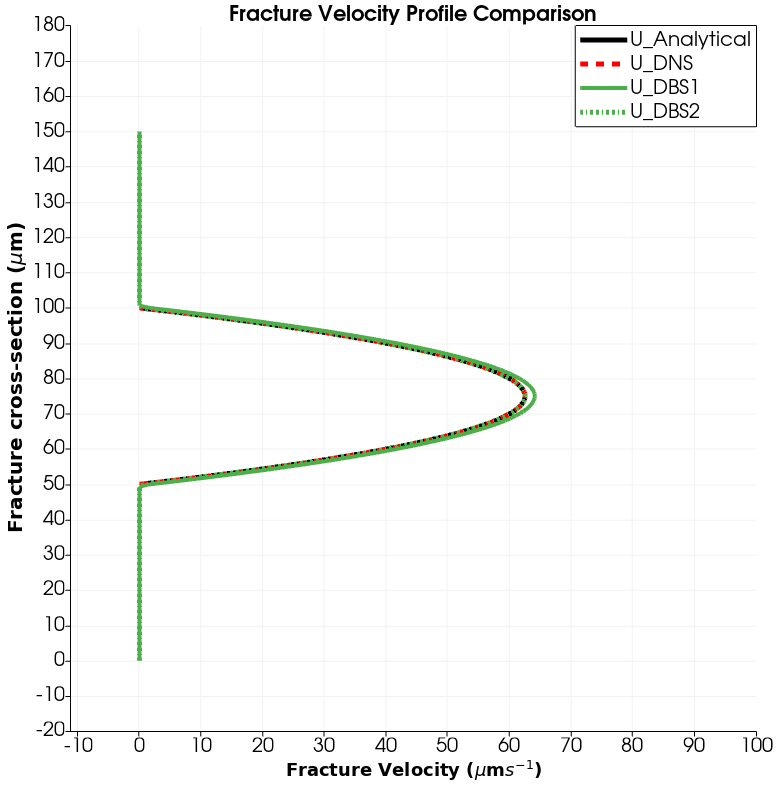}
  \caption{Velocity profile in the microfracture at a mesh resolution of 0.625$\mu$m. }
  \label{fig:velocityProfile}
\end{figure}

A quantitative comparison of the velocity prediction accuracy is provided in Table~\ref{tab:L2_error_channelDBS}. The $L_2$ (root-mean-square, RMS) error, defined as the root-mean-square of the velocity difference normalised by the domain volume, serves as a measure of the overall magnitude of the prediction error and is given by
\begin{equation}
L_2 = \left( \frac{1}{V} \int_V \| \mathbf{u} - \mathbf{u}_{\mathrm{ref}} \|^2 \, dV \right)^{1/2},
\end{equation}
where $\mathbf{u}$ is the predicted velocity field, $\mathbf{u}_{\mathrm{ref}}$ is the reference velocity field, and $V$ is the total domain volume. The computed $L_2$  (RMS) error in the 0.625$\mu$m resolution for DBS1 is $1.56 \times 10^{-5}$, whereas the corresponding value for DBS2 is effectively zero. These results confirm that the DBS2 approach reproduces the DNS velocity field with significantly higher accuracy than the DBS1 method. These findings are further evident in the velocity profile plot (Figure \ref{fig:velocityProfile}) for the different methods at a mesh resolution of 0.625$\mu$m.

\begin{table}[ht!]
\centering
\caption{L2 (RMS) error comparison for DBS1 and DBS2 methods.}
\begin{tabular}{l c c}
\hline
\textbf{Method} & \textbf{DBS1} & \textbf{DBS2} \\
\hline
L2 (RMS) Error & 1.56 $\times 10^{-5}$  & $1 \times 10^{-21}$ \\
\hline
\end{tabular}
\label{tab:L2_error_channelDBS}
\end{table}

\begin{figure}[H]
  \centering
  \includegraphics[width=0.6\textwidth]{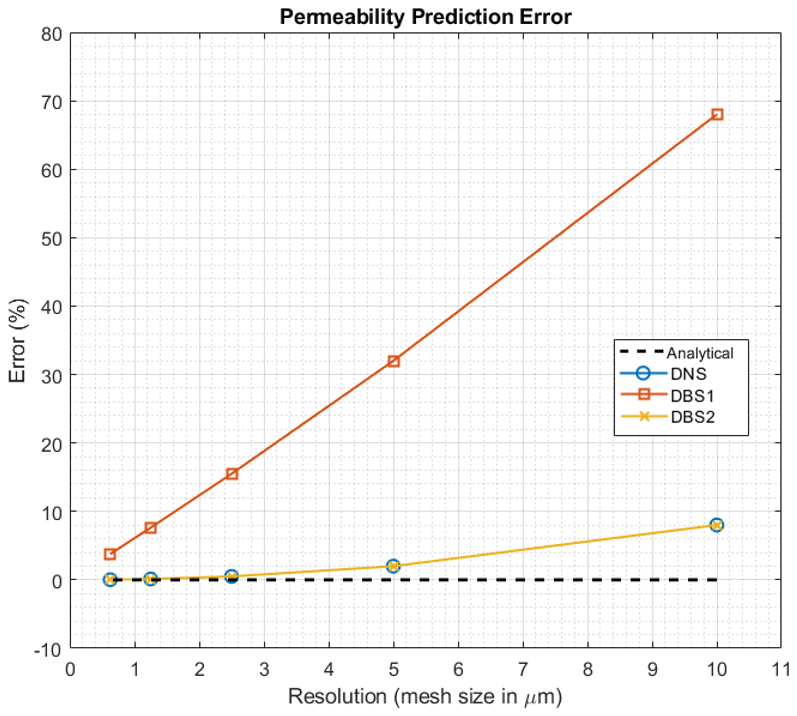}
  \caption{Convergence rate of predicted permeability of the 2D microchannel for different mesh sizes.}
  \label{fig:permeabilityPrediction}
\end{figure}

Figure~\ref{fig:permeabilityPrediction} shows the predicted permeability of the 2D microchannel from numerical simulations using different methods at varying mesh resolutions. The simulation results are plotted against the analytical solution. The results show that the accuracy of all methods depends on mesh resolution and that they converge to the analytical solution as the mesh is refined. The DBS2 method converges at a rate nearly identical to that of the DNS method, whereas the DBS1 method converges poorly. The data in Table~\ref{tab:dbs_error} show the relative error (in $\%$) for each method at five successive mesh refinements.

\begin{table}[ht!]
\centering
\caption{Relative error in predicted permeability for DNS, DBS1, and DBS2 at different mesh resolutions.}
\begin{tabular}{c c c c}
\hline
\multirow{2}{*}{Mesh Resolution ($\mu$m)} & \multicolumn{3}{c}{Relative Error in Predicted Permeability ($\%$)} \\
\cline{2-4}
 & DNS & DBS1 & DBS2 \\
\hline
10   & 8.01   & 68.0  & 8.01 \\
5   & 2.00   & 32.0  & 2.00 \\
2.5   & 0.501  & 15.5  & 0.501 \\
1.25   & 0.126  & 7.63  & 0.126 \\
0.625   & 0.031  & 3.78  & 0.031 \\
\hline
\end{tabular}
\label{tab:dbs_error}
\end{table}

Similar to the observation in Figure~\ref{fig:permeabilityPrediction}, the DNS and DBS2 display the same rapid convergence with mesh refinement. The relative error for both methods falls from 8.01$\%$ (coarsest mesh) to 0.031$\%$ (finest mesh). At each halving of the mesh size, the error is reduced by roughly a factor of 4. At the finest mesh, DBS1 has a 3.78$\%$ error while DNS and DBS2 have 0.031$\%$.
This further demonstrates that accounting for the second Brinkman term in our DBS2 formulation significantly improves the method's predictive accuracy. 

\subsection {Two-Dimensional (2D) Micromodel} \label{subsec:2D_micromodel}
\begin{figure}[H]
  \centering
  \includegraphics[width=0.6\textwidth]{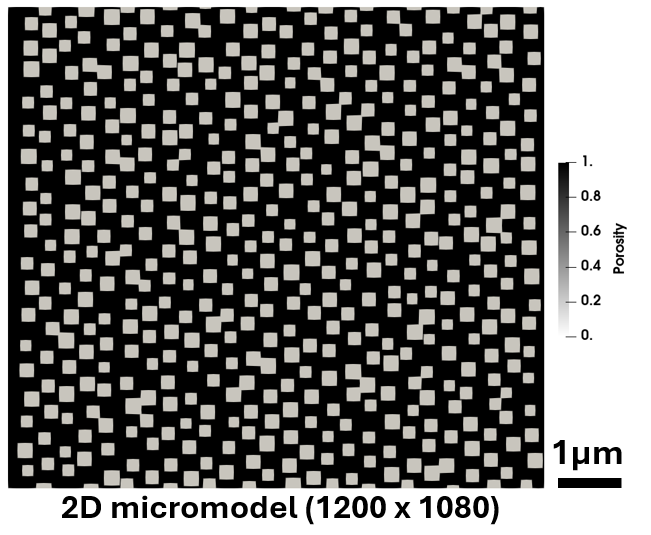}
  \caption{2D heterogeneous micromodel image with a dimension of 1200 x 1080 (gray = solid beads; black = pores; resolution of $1~\mu m$ - not to scale).}
  \label{fig:micromodel}
\end{figure}

Another validation case considers a 2D micromodel populated with randomly distributed square beads of varying sizes. The micromodel measures $1200 \times 1200$ voxels at a spatial resolution of $1~\mu$m. Grey voxels represent the solid phase (beads), while black voxels denote the pore space, as illustrated in Figure~\ref{fig:micromodel}. The micromodel geometry is generated synthetically by randomly placing square beads with prescribed size distributions within the domain, thereby yielding controlled yet heterogeneous pore structures representative of experimental micromodels. The creation of this micromodel is based on the open-source repository publicly available at  
\url{https://github.com/hannahmenke/DrawMicromodels}.

To mitigate boundary effects and avoid preferential flow paths at the top and bottom of the geometry, the model is cropped to 1200 × 1080. For the DBS simulations, the computational mesh is generated using the blockMesh utility, consistent with the previous case \ref{subsec:2D_Fracture}. For the DNS simulations, the OpenFOAM utilities topoSet and subsetMesh are used in conjunction with blockMesh to generate meshes that accurately represent the pore geometry. This workflow enables the extraction of identical pore configurations for both the DNS and DBS cases. As a result, all simulations are performed on geometrically consistent domains, providing a fair and controlled basis for comparison.

\subsubsection{Numerical Simulation Setup}

The flow of a fluid with viscosity $1 \times 10^{-6}~\mathrm{m^2/s}$ is simulated within the micromodel. A kinematic pressure of $0.1~\mathrm{m^2/s^2}$ is imposed at the inlet and $0$ at the outlet, generating a pressure drop of $0.1~\mathrm{m^2/s^2}$ across the domain. Zero-gradient boundary conditions are applied to the velocity at both the inlet and the outlet, allowing the flow field to adjust naturally to the imposed pressure difference. A summary of the boundary conditions is provided in Table~\ref{tab:bc_micromodel}.

\begin{table}[H]
\centering
\caption{Boundary conditions for the micromodel test case with square beads.}
\begin{tabular}{l l l}
\hline
\textbf{Boundary} & \textbf{Field} & \textbf{Boundary condition} \\
\hline
Inlet  & Pressure & fixedValue (\(p_{\mathrm{in}}\)) \\
       & Velocity & zeroGradient \\
Outlet & Pressure & fixedValue (\(p_{\mathrm{out}}\)) \\
       & Velocity & zeroGradient \\
Top \& Bottom & All fields & symmetryPlane (symmetry) \\
Front \& Back & All fields & Empty \\
\hline
\end{tabular}
\label{tab:bc_micromodel}
\end{table}

\noindent
Using symmetry boundary conditions on the top and bottom surfaces minimises artificial boundary layers induced by the channel walls and more closely approximates an unbounded porous medium in the vertical direction. 

\subsubsection*{Results and Discussions}

The predicted permeabilities for each method are summarised in Table~\ref{tab:micromodel_perm}. DNS yields a permeability of \(1.493 \times 10^{-11}~\text{m}^2\). DBS2 closely reproduces this value at \(1.515 \times 10^{-11}~\text{m}^2\), corresponding to a relative error of only \(1.5\%\). By contrast, DBS1 significantly overpredicts the permeability (\(1.703 \times 10^{-11}~\text{m}^2\)) with a relative error of \(14.1\%\). This improvement highlights the role of the second Brinkman term in better capturing interfacial viscous dissipation, which is otherwise underestimated in DBS1. Nevertheless, the accuracy of DBS2 in this heterogeneous 2D micromodel is lower than in the microfracture test case \ref{subsec:2D_Fracture}, due to the increased tortuosity of the system, which induces non-linear pressure fields around the solid beads.

\begin{table}[H]
\centering
\caption{Permeability predictions and relative error for the 2D micromodel.}
\label{tab:micromodel_perm}
\begin{tabular}{l c c c}
\hline
Property & DNS & DBS1 & DBS2 \\
\hline
Permeability (\( \times 10^{-11}~\text{m}^2\)) & 1.493 & 1.703 & 1.515 \\
Error (\%) & -- & 14.1 & 1.5 \\
\hline
\end{tabular}
\end{table}

\begin{figure}[H]
  \centering
  \includegraphics[width=1\textwidth]{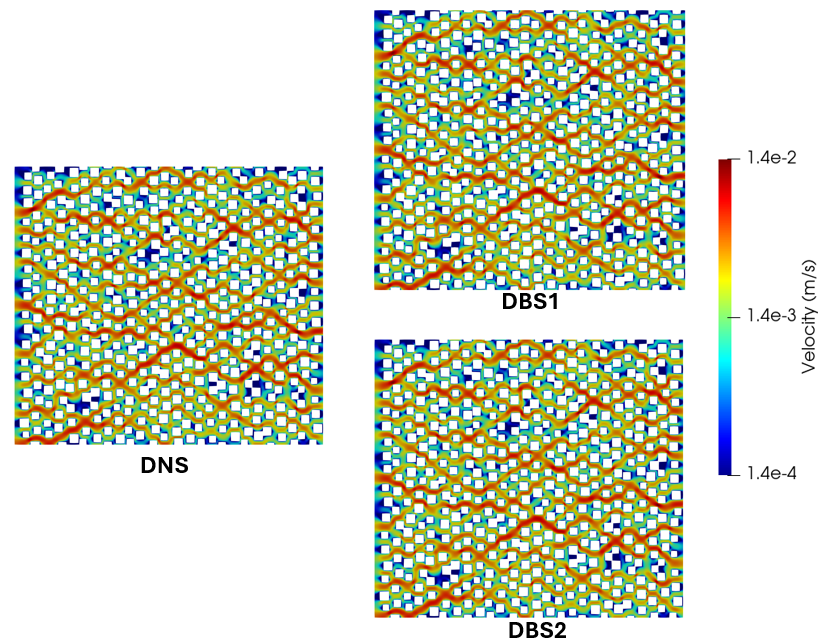}
  \caption{Simulation result showing the velocity field distribution in the micromodel for the different methods.}
  \label{fig:micromodelResults}
\end{figure}

Velocity field maps (Figure~\ref{fig:micromodelResults}) show that all three approaches predict very similar preferential flow paths through the pore space, and the global range of velocities is comparable (DNS/DBS2 reaching $\max |U|\approx 0.013~\mathrm{m/s}$, while in DBS1 $\max |U| \approx 0.014~\mathrm{m/s}$). However, the key difference is evident in small over-predictions of velocity occurring across much of the pore space (particularly around constrictions and near solid boundaries), consistent with an underestimation of viscous resistance in the standard DBS formulation.

\begin{figure}[h!]
  \centering
  \includegraphics[width=0.8\textwidth]{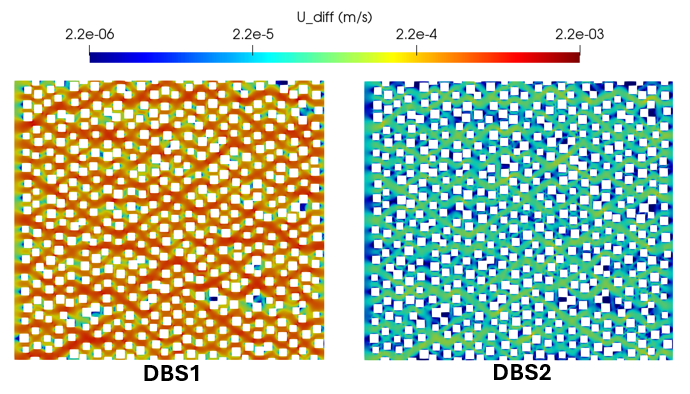}
  \caption{Velocity difference map for DBS1 and DBS2 as compared to the DNS method.}
  \label{fig:micromodel_diff_map}
\end{figure}

The cumulative effect of these velocity errors is substantial, producing about 14\% overestimation of permeability (Table~\ref{tab:micromodel_perm}). By contrast, DBS2 reduces these distributed errors, bringing both velocity and permeability predictions closer to DNS results. This behaviour is consistent with the horizontal fracture benchmark (Figure~\ref{fig:permeabilityPrediction}), in which DBS2 converges rapidly to DNS with mesh refinement, whereas DBS1 remains offset. Overall, these results indicate that the second Brinkman term primarily corrects distributed viscous drag, thereby improving integral quantities (e.g., permeability) even when the qualitative velocity field appears similar. Figure~\ref{fig:micromodel_diff_map} presents the velocity difference ($U_{\text{diff}}$) maps for the DBS1 and DBS2 methods, compared with the reference DNS results in the micromodel. In contrast to the fracture case, where the DBS2 method reproduced the DNS velocity field with perfect accuracy, the micromodel results reveal significant deviations for both methods. The heterogeneous pore geometry produces spatially varying pressure and velocity fields. Residual differences from DNS remain locally for both DBS formulations, particularly near constrictions and fluid--solid interfaces, with DBS2 showing closer agreement with DNS.

Quantitative assessment through $L_2$ (RMS) error analysis (Table \ref{tab:l2_error_micromodel}) confirms these observations. Using this metric, DBS1 exhibits an error of $33.8 \times 10^{-5}$, whereas DBS2 achieves a substantially reduced error of $4.36 \times 10^{-5}$.
 
\begin{table}[ht!]
\centering
\caption{L2 (RMS) error comparison for DBS1 and DBS2 methods.}
\begin{tabular}{l c c}
\hline
\textbf{Method} & \textbf{DBS1} & \textbf{DBS2} \\
\hline
L2 (RMS) Error ($\times 10^{-5}$) & 33.8 & 4.36 \\
\hline
\end{tabular}
\label{tab:l2_error_micromodel}
\end{table}

This corresponds to an approximately $87\%$ reduction in the RMS velocity error, demonstrating improved agreement with DNS for DBS2, even though both DBS1 and DBS2 methods fail to capture the complex nonlinear flow dynamics in tortuous pore geometries.
  
\subsection {Three Dimensional (3D) Bentheimer Image}
As a final validation case, we consider fluid flow through a 3D micro-CT image of a Bentheimer sandstone sample, which provides a more realistic geometry. The image has dimensions \(400^{3}\) voxels with a spatial resolution of \(5~\mu\text{m}\). As is common in DRP, the micro-CT data are provided as binary images, where each voxel represents either solid or pore. 

\subsubsection{Mesh generation}
\begin{figure}[!h]
  \centering
  \includegraphics[width=1\textwidth]{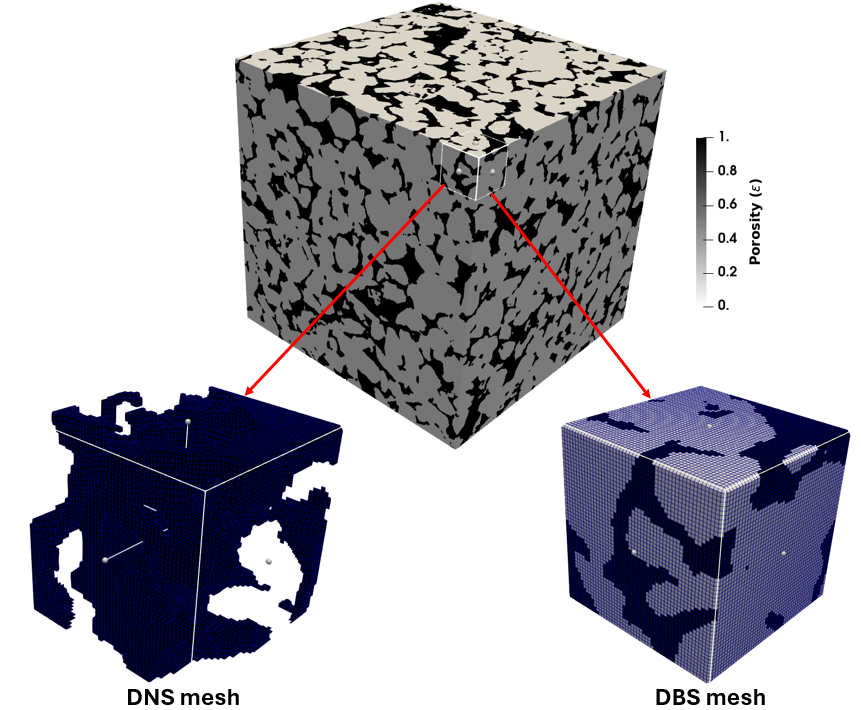}
  \caption{An image of a  \(400^{3}\) Bentheimer sample showing the meshing configuration of the DNS and DBS methods.}
  \label{fig:BentheimerandMesh}
\end{figure}

For the simulations, the computational domains were initialised as voxelised Cartesian meshes, in which each voxel corresponds to a cubic cell classified as either a pore or a solid. The voxelized meshes were generated using the \texttt{blockMesh} utility, after which porosity values were assigned to each voxel, with \(1 \times 10^{-4}\) for solid regions and a value of 1 for pore regions. A workflow similar to that described in Section \ref{subsec:2D_micromodel} was applied for the DNS cases, in which pore-space meshes were extracted from the voxelized domains using the \texttt{topoSet} and \texttt{subsetMesh} utilities in OpenFOAM, which were configured to remove solid voxels and retain only the connected pore-space (see Figure \ref{fig:BentheimerandMesh}). This procedure ensured that the DNS pore-space mesh was consistent with the geometry used in the DBS simulations, allowing for a direct comparison between the two approaches. In contrast, the DBS simulations employed the full voxelized domains, in which both solid and pore voxels were retained, consistent with the micro-continuum formulation of the DBS method.

\subsubsection*{Boundary Conditions}
A kinematic pressure of $1~\mathrm{m^2/s^2}$ is imposed at the inlet and $0~\mathrm{m^2/s^2}$ at the outlet, giving a kinematic pressure drop of $1~\mathrm{m^2/s^2}$ across the Bentheimer sample. The boundary conditions
are summarised in Table~\ref{tab:bc_Bentheimer}. The fluid kinematic viscosity is $\nu=1\times10^{-6}~\mathrm{m^2/s}$.

\begin{table}[!h]
\centering
\caption{Boundary conditions for the Bentheimer test case.}
\begin{tabular}{l l l}
\hline
\textbf{Boundary} & \textbf{Field} & \textbf{Boundary condition} \\
\hline
Inlet  & Pressure & fixedValue (\(p_{\mathrm{in}}\)) \\
       & Velocity & zeroGradient \\
Outlet & Pressure & fixedValue (\(p_{\mathrm{out}}\)) \\
       & Velocity & zeroGradient \\
Walls & Velocity & noSlip \\
\hline
\end{tabular}
\label{tab:bc_Bentheimer}
\end{table}

\subsubsection*{Results and Discussions}
The Bentheimer sandstone results demonstrate a marked difference in performance between methods compared to the simpler test cases. The permeability predictions are shown in Table~\ref{tab:bentheimer_perm}, where DNS establishes a reference value of (\(5.513 \times 10^{-12}\text{m}^2\)). DBS1 substantially overestimates this at $(7.602 \times 10^{-12}\text{m}^2)$ (37.9$\%$ error), while DBS2 achieves closer agreement at $(5.722 \times 10^{-12}~\text{m}^2)$ (3.8\% error).

\begin{table}[ht!]
\centering
\caption{Permeability predictions and relative error for the 3D Bentheimer sandstone.}
\label{tab:bentheimer_perm}
\begin{tabular}{l c c c}
\hline
Property & DNS & DBS1 & DBS2 \\
\hline
Permeability (\( \times 10^{-12}~\text{m}^2\)) & 5.513 & 7.602 & 5.722 \\
Error (\%) & -- & 37.9 & 3.8 \\
\hline
\end{tabular}
\end{table}
The large error in DBS1 is particularly striking when compared to its performance in previous cases. This suggests that the standard Brinkman formulation struggles with the geometric complexity inherent in natural rock samples. The tortuous pore geometry, sharp porosity gradients, and varying pore throat sizes in the Bentheimer case create conditions under which accurate viscous dissipation modelling is critical and this why our proposed DBS2 formulation performs better in this example.

\begin{figure}[!h]
  \centering
  \includegraphics[width=1.05\textwidth]{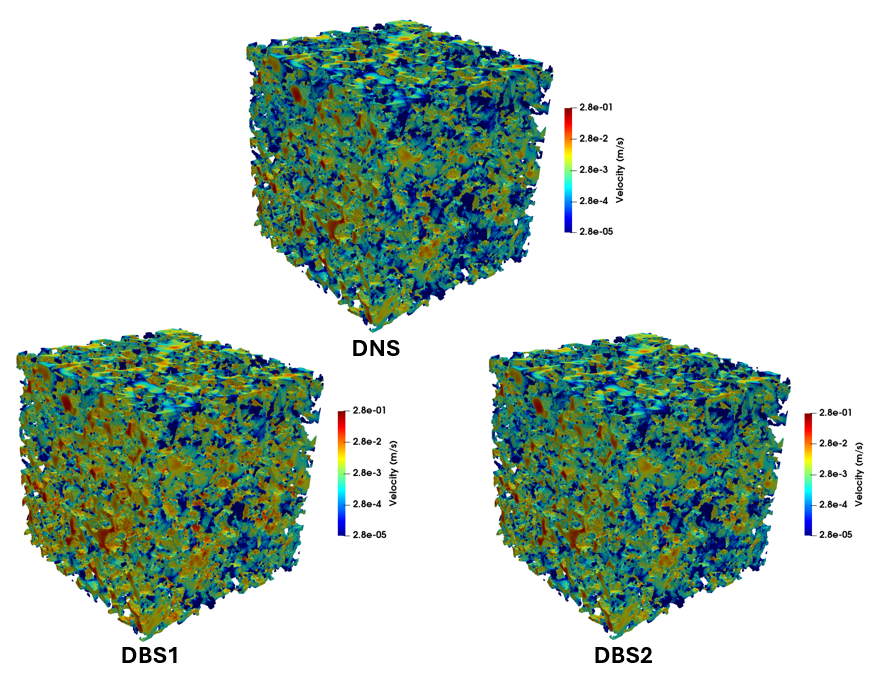}
  \caption{Velocity field maps for the  \(400^{3}\) Bentheimer sample comparing the different methods.}
  \label{fig:BentheimerResults}
\end{figure}

\begin{figure}[!h]
  \centering
  \includegraphics[width=1\textwidth]{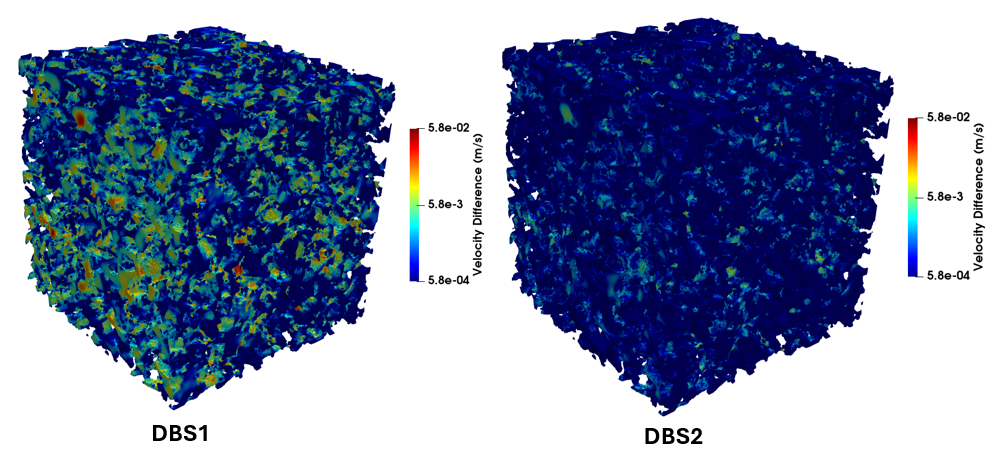}
  \caption{Velocity difference map for the  \(400^{3}\) Bentheimer sample comparing the velocity error between DBS1 and DBS2 relative to DNS.}
  \label{fig:Bentheimerudifference}
\end{figure}

Examining the velocity fields (Figure~\ref{fig:BentheimerResults}) reveals apparent differences in pore-scale flow. DNS produces the expected heterogeneous patterns, characterised by preferential pathways. DBS2 reproduces these features with high accuracy, capturing both the dominant flow channels and their associated tortuosity and velocity distributions. In contrast, DBS1, while showing similar features, exhibits markedly different behaviour, failing to resolve the viscous boundary effects that control resistance in the narrow pore throats typical of Bentheimer sandstone. This results in an artificial enhancement of flow capacity and a systematic overestimation of flow velocity and permeability. 
The difference between the computed velocity fields of DBS1 and DBS2 and the DNS is shown in Figure~\ref{fig:Bentheimerudifference} and further supports these arguments. The complexity of the three-dimensional pore structure amplifies these discrepancies. Unlike the controlled geometries of earlier test cases, Bentheimer contains a challenging combination of variable pore sizes, irregular shapes, and complex connectivity. Under such conditions, inclusion of the second Brinkman term in DBS2 is essential, as it corrects for viscous losses that DBS1 underestimates. These observations have direct implications for digital rock analysis, where computational efficiency must be balanced against the need for accuracy and precision. The 37.9$\%$ error associated with DBS1 is unacceptable for most engineering applications, whereas the 3.8$\%$ error of DBS2 lies within typical experimental uncertainty. Moreover, DBS2 retains significant computational advantages over DNS because it operates directly on voxelized domains, eliminating the need for additional mesh manipulation for pore-space extraction, making it well-suited for routine analysis of micro-CT rock samples.

\section{Conclusions}

This study presents a comprehensive validation of a conservative Darcy-Brinkman–Stokes (DBS) formulation for pore-scale flow simulations, demonstrating that the full viscous contribution of the DBS formulation can be recovered without introducing a separate discretisation of the second Brinkman term (SBT). Across validation cases of increasing complexity—ranging from idealised microchannels/fractures to synthetic micromodels and natural Bentheimer sandstone—the DBS2 formulation consistently improves agreement with direct numerical simulation (DNS), while the classical formulation (DBS1) systematically overpredicts permeability. These results demonstrate that the conservative DBS2 formulation accurately represents viscous boundary effects in heterogeneous porous media, without requiring additional numerical treatment beyond the standard finite-volume discretisation.

From a numerical perspective, a key strength of the implementation is its expression of the viscous operator in conservative form. This formulation allows the SBT to be incorporated naturally through a variable-coefficient diffusion term, without introducing additional source terms or complex discretisation of porosity gradients. As a result, both the classical Brinkman diffusion and SBT are captured within a single finite-volume operator, thereby avoiding explicit storage of gradients while preserving the robustness and efficiency of the implicit framework. The conservative formulation is therefore central to enabling accurate, computationally efficient representation of higher-order viscous effects in pore-scale micro-continuum simulations. The formulation is implemented within the open-source GeoChemFoam framework.

Beyond the study’s fundamental academic relevance, its implications extend directly to Digital Rock Physics, where accurate yet computationally affordable pore-scale simulations are essential for the routine analysis of micro-CT image datasets. The DBS micro-continuum framework, enabled by a conservative formulation of the viscous operator, provides a practical pathway to bridge pore-scale physics with continuum-scale models. This capability supports a wide range of subsurface applications, including carbon and hydrogen storage and geothermal energy recovery. 

Future work could extend the current validation to multiphase and reactive-flow systems, where additional physical coupling may impose stronger demands on the conservative DBS formulation. Although the DBS2 formulation demonstrates strong accuracy and efficiency, further work is needed to assess its behaviour across a wider range of pore geometries, particularly highly tortuous ones, where nonlinear pressure variations and complex local flow structures may not be fully captured at a given voxel resolution. Addressing these challenges would further improve predictive accuracy and broaden the applicability of the DBS approach in Digital Rock Physics.

%\section*{References}
\printbibliography

\section*{Statements and Declarations}

\subsection*{Funding}
Gospel E. Stewart acknowledges financial support from the Petroleum Technology Development Fund (PTDF).

\subsection*{Competing Interests}
The authors have no relevant financial or non-financial interests to disclose.

\subsection*{Author Contributions}
Gospel E. Stewart developed the numerical implementation, performed the simulations, analysed the results, and wrote the first draft of the manuscript. Hannah P. Menke, David Egya, Jacqueline Mifsud, and Julien Maes contributed to the conceptual development, interpretation of results, supervision, and revision of the manuscript. All authors read and approved the final manuscript.

\subsection*{Data Availability}
The test cases used in this study are publicly available in the GeoChemFoam repository at \url{https://github.com/GeoChemFoam/GeoChemFoam/tree/main/Examples/ConservativeDBSPaper}.

\end{document}